\documentclass[acmsmall]{acmart}

\AtBeginDocument{%
  \providecommand\BibTeX{{%
    \normalfont B\kern-0.5em{\scshape i\kern-0.25em b}\kern-0.8em\TeX}}}

\setcopyright{acmcopyright}
\copyrightyear{2018}
\acmYear{2018}
\acmDOI{10.1145/1122445.1122456}

\acmJournal{CSUR}
\acmVolume{37}
\acmNumber{4}
\acmArticle{}
\acmMonth{8}

\begin{document}

%
\title{AI-Empowered Persuasive Video Generation: A Survey}

%
\author{Chang Liu}
\email{chang015@e.ntu.edu.sg}
\orcid{1234-5678-9012}
\author{Han Yu}
\email{han.yu@ntu.edu.sg}
\affiliation{%
  \institution{School of Computer Science and Engineering, Nanyang Technological University}
  \country{Singapore}
}

%
\renewcommand{\shortauthors}{Liu and Yu, et al.}

%
\begin{abstract}
Promotional videos are rapidly becoming a popular medium for persuading people to change their behaviours in many settings (e.g., online shopping, social enterprise initiatives). Today, such videos are often produced by professionals, which is a time-, labour- and cost-intensive undertaking. In order to produce such contents to support a large applications (e.g., e-commerce), the field of artificial intelligence (AI)-empowered persuasive video generation (AIPVG) has gained traction in recent years. This field is interdisciplinary in nature, which makes it challenging for new researchers to grasp. Currently, there is no comprehensive survey of AIPVG available. In this paper, we bridge this gap by reviewing key AI techniques that can be utilized to automatically generate persuasive videos. We offer a first-of-its-kind taxonomy which divides AIPVG into three major steps: 1) visual material understanding, which extracts information from the visual materials (VMs) relevant to the target of promotion; 2) visual storyline generation, which shortlists and arranges high-quality VMs into a sequence in order to compose a storyline with persuasive power; and 3) post-production, which involves background music generation and still image animation to enhance viewing experience. We also introduce the evaluation metrics and datasets commonly adopted in the field of AIPVG. We analyze the advantages and disadvantages of the existing works belonging to the above-mentioned steps, and discuss interesting potential future research directions.
\end{abstract}

%
%


%

%

%
\maketitle

\section{Introduction}

In e-commerce platforms (e.g., Taobao, Lazada, JD), promotional videos are rapidly becoming a popular form of product advertising. According to reports\footnote{\url{https://www.wyzowl.com/sovm-results-2020/}}, 80\% of video marketeers stated that videos had helped increase their sales. The production of promotional videos has hitherto been performed mainly by professional teams. For individual or micro e-commerce sellers, hiring a professional video production team is unaffordable. Besides, manual video production is time-consuming, which hinders large-scale adoption of video-based product promotion campaigns.

To address this issue, artificial intelligence (AI) techniques for generating videos have been developed \cite{hua_photo2videosystem_2006,chen_tiling_2006,choi_video-story_2016,liu2020ai,liu2020generating}. This field can be referred to as AI-empowered persuasive video generation (AIPVG). AIPVG systems usually take visual materials (VMs) (e.g., images and video clips) provided by sellers as the input, and generate videos as the output. One of the main design objectives of AIPVG is to motivate the viewers to change their behaviours (e.g., purchase products from an e-commerce platform, adopt an active and healthy lifestyle). Thus, generating videos that are persuasive is a major goal of AIPVG.

Persuasion has been a subject of study in social sciences for decades \cite{andersen1971persuasion,perloff1993dynamics}. \citeauthor{perloff1993dynamics} \cite{perloff1993dynamics} defines persuasion as a symbolic process in which communicators try to convince other people to change their attitudes or behaviours regarding an issue through the transmission of a message, in an atmosphere of free choice. In  e-commerce, the transmission of messages can be carried out via images, texts and promotional videos. The goal is to influence potential buyers' purchasing decisions. Here, we provide a brief overview of persuasion models commonly adopted by AIPVG literature.

Elaboration Likelihood Model (ELM) \cite{petty_elaboration_1986} divides the process of persuasion into two routes: 1) the central route, and 2) the peripheral route. Persuasion through the central route requires the persuader to carefully analyse the facts surrounding the subject of persuasion in order to formulate a persuasive message. On the other hand, the peripheral route relies on auxiliary cues of the persuasion context (e.g., aesthetics, building relationships with the persuadees). In \cite{petty_elaboration_1986}, the authors proposed to use peripheral routes when the persuadee has no motivation to analyze, cannot process information, or holds a neutral attitude towards the issue. Note that if a persuadee holds unfavourable thoughts, the peripheral routes may not be able to change his/her mind.

Focusing more on the knowledge of both the persuadee and persuader, the Persuasion Knowledge Model (PKM) \cite{friestad1994persuasion} models the persuasion process in an adversarial form. The persuader performs persuasion based on his/her knowledge, while the persuadee copes with the persuasion attempt. Three types of knowledge are considered in the PKM: 1) topic knowledge, which refers to the knowledge related to the subject of persuasion (e.g., a product); 2) persuasion knowledge, which is about how to successfully persuade people (for persuaders), or how to cope with such persuasion (for persuadees); and 3) persuadee/persuader knowledge, which is the information that the persuadee or the persuader know about each other.

Applying AI techniques to videos generation has been a long-running research field. Many survey papers have been published on this broad topic. For example, \textit{video summarization} techniques have been surveyed in  \cite{hussain2021comprehensive,apostolidis2021video,sebastian2015survey}. Though both video summarization and AIPVG generate videos that can tell stories, the design objectives and the inputs of these techniques are very different. Specifically, the video summarization trims a long video into a short version while aiming to preserve the original information. AIPVG, on the other hand, takes a set of VMs as the input to generate videos that aim to persuade viewers to change certain aspects of their behaviour. Several other papers have surveyed the field of \textit{video synthesis with generative models} \cite{chen2020comprises,jong2020virtual}. Such techniques do not require a set of VMs as the input to generate the videos. However, they can only generate videos without sophisticated storylines (e.g., talking-head videos \cite{chen2020comprises} or virtual try-on videos \cite{jong2020virtual}). AIPVG generates videos with storylines designed for persuasion by selecting and sequencing input VMs. This line of research works is distinct from the topics covered by the aforementioned survey papers. Currently, to the best of our knowledge, there is no comprehensive survey on the topic of AIPVG.

The techniques involved in AIPVG are interdisciplinary in nature, involving computer science, social sciences and film theory. Thus, it can be challenging for new-comers to grasp the important developments in this field. The lack of a comprehensive survey on this topic further exacerbates this challenge. In this paper, we bridge this gap by providing a comprehensive review of the existing literature on AIPVG. This paper contributes to the AI literature in the following ways:
\begin{itemize}
    \item Firstly, we provide an overview of the persuasion theory which plays an important role in AIPVG approach designs. This can be helpful for researchers to gain useful background knowledge on the ways through which technologies can influence users' behaviours.
    \item Secondly, we propose a unique taxonomy which divides existing works in AIPVG into three main categories: 1) visual material understanding, 2) visual storyline generation, and 3) post-production. For each of these categories, we provide an overview of the main research challenges, discuss the key ideas of notable works, and highlight potential areas for improvement. We also introduce how to evaluate the generated videos.
    \item Finally, we outline ten promising future research directions towards building practical and efficient AIPVG approaches.
\end{itemize}




\section{The Proposed AIPVG Taxonomy}
\begin{figure}[ht]
    \centering
    \includegraphics[width=1.02\linewidth]{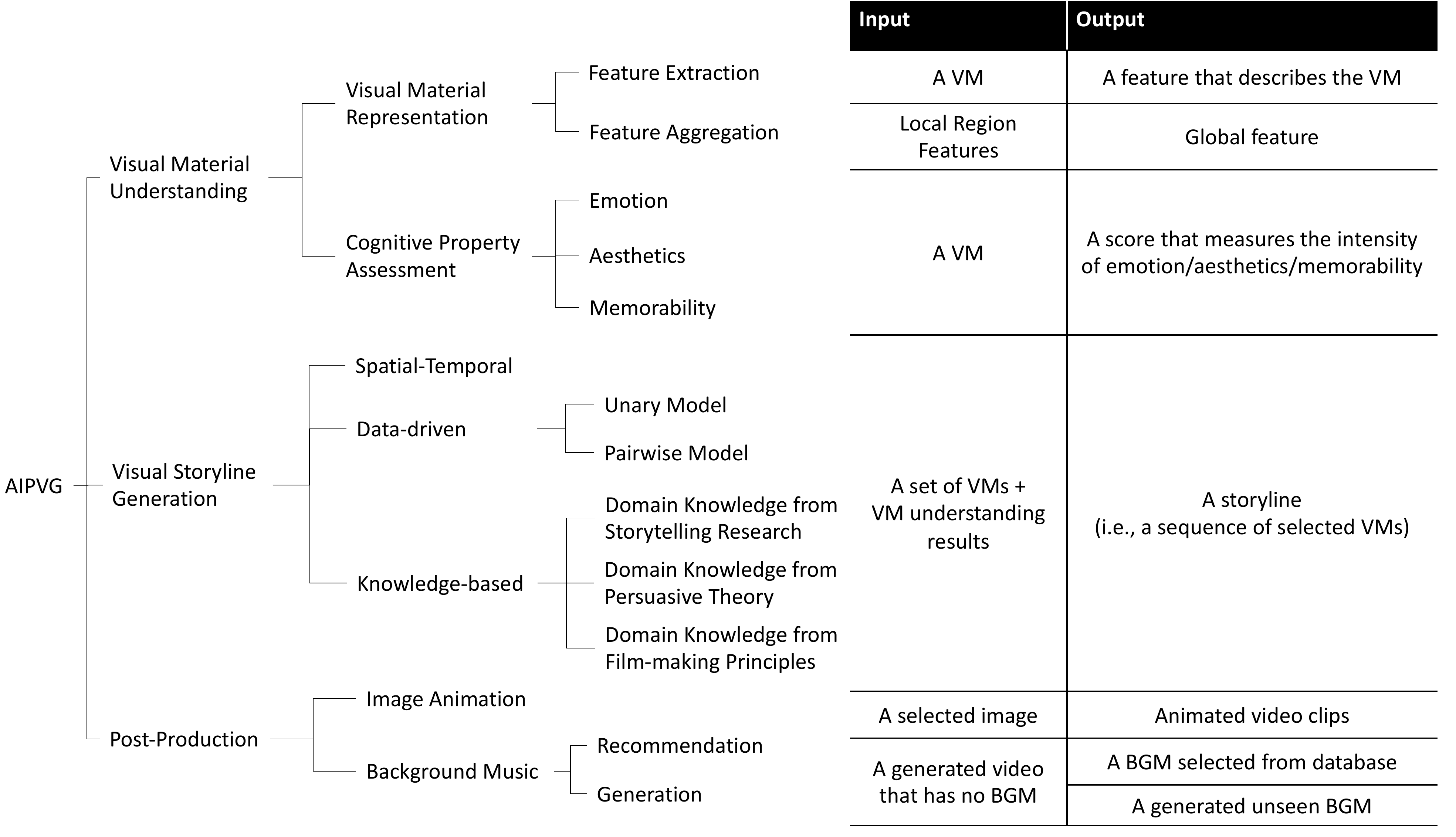}
    \caption{The proposed taxonomy of the AIPVG literature.}
    \label{fig:taxonomy}
\end{figure}
The proposed AIPVG taxonomy divides existing works in this field into three main categories, as shown in Figure~\ref{fig:taxonomy}. Each of these categories correspond to one important step in the process of generating promotional videos with AI: 
\begin{enumerate}
    \item \textbf{Visual Material Understanding}: Works in this category aim to enable AI to extract useful information from images or videos provided by the sellers relevant to the target of promotion. They generally serve to enhance the informativeness and attractiveness of the generated promotional videos, which can be utilized by the subsequent steps.
    \item \textbf{Visual Storyline Generation}: Works in this category focus on selecting and ordering the visual materials to compose a persuasiveness sequence. In this step, the informativeness and attractiveness of the visual materials are often taken into account to facilitate the selection and sequencing operations.
    \item \textbf{Post-Production}: Works in this category aim to improve the viewing experience for the users. They include background music generation, image animation, and video rendering with special effects. This step can add more style and attractiveness to the final video output.
\end{enumerate}

In the following sections, we discuss the main research challenges, ideas of notable works, and highlight potential areas for improvement in each of the aforementioned categories.

\section{Visual Material Understanding}
Visual Material Understanding techniques that are useful for video generation can be divided into two parts: 1) Visual Material Representation, which takes an VM as the input and extracts a feature from the VM. A good representation preserves information about the target object depicted in the image. It serves as a \textit{low-level understanding} of the VMs for video generation and can be helpful to facilitate informativeness based central route of persuasion \cite{petty_elaboration_1986}; and 2) Cognitive Property Assessment (CPA), which estimates scores of cognitive properties for given VMs. It aims to predict the intensity of stimulation when viewers watch the VMs. Thus, it is considered as a \textit{high-level understanding} approach. The cognitive properties help identify high quality VMs for the peripheral route of persuasion \cite{petty_elaboration_1986}. Commonly adopted cognitive properties in visual material understanding include emotions, aesthetics and memorability.

\subsection{Visual Material Representation}
To generate promotional videos by leveraging AI techniques, the first step is to represent the visual materials (VMs). The VM representations can be used to perform tasks important for VM sequence generation and post-production (e.g., clustering, similarity calculation and saliency map generation).
A high quality VM representations generation model needs to satisfy the following requirements: 1) it shall generate embeddings that can be easily processed by AI models (e.g, 1-D vectors); and 2) it shall generate embeddings which preserve the rich product information, which is crucial in enhancing the informativeness of the resulting visual storylines, thus improving their persuasiveness \cite{armstrongPersuasiveAdvertisingEvidencebased2010}. 

\subsubsection{Feature Extraction}
Feature extraction approaches for visual material understanding have evolved over the years from traditional methods, to deep learning classification methods, and more recently to deep metric learning-based methods.

\textbf{Traditional Methods:}
Before the emergence of Convolutional Neural Networks, feature engineering was generally performed manually by experts. SIFT and SURF \cite{lowe1999object,bay2006surf} have been commonly used to extract image features. The scale-invariant feature transform (SIFT) \cite{lowe1999object} uses local features near detected key points of an image as the representation. The local features are calculated based on the gradients of the pixels in each sub-region. Rotation and scale normalization are then performed on the features. SURF further leverages an integral image to accelerate SIFT \cite{bay2006surf}. 
These traditional image representation methods are generally fast. However, the major drawback is that they cannot learn what features to be extracted from the input images. It heavily relies on manual feature engineering, which is hard to scale up. The model performance is also affected by the feature engineers' expertise levels. 

\textbf{Classification-based Deep Learning Methods:} \label{sec:cls}
Classification-based deep learning VM representation generation methods use the middle layer outputs of a trained classification Deep Neural Networks (DNN) as the feature. Commonly adopted classification DNNs for this task include convolutional neural networks (CNNs) such as AlexNet \cite{krizhevsky2014}, ResNet \cite{he_deep_2016} and VGG \cite{simonyan_very_2015}. Recent research also suggests that the transformer networks \cite{dosovitskiy2020image} can achieve superior performance in classification. 3D-CNNs \cite{feichtenhofer2019slowfast} can be used for video classification and feature extraction. Classification DNNs are trained by minimizing the classification loss via gradient descent optimization (e.g., stochastic gradient descent (SGD), Nesterov SGD \cite{sutskever2013importance} and Adam \cite{kingma2014adam}). Commonly used loss functions are the cross-entropy loss and focal loss \cite{lin2017focal}. 
In practice, to extract the features in general, it is common to use the DNNs which have been pre-trained on well-known large-scale datasets (e.g., ImageNet \cite{deng_imagenet:_2009} for images, and Kinetics-400 \cite{kay2017kinetics} for videos) as the feature extractors. Since such pre-trained DNNs are often available online, the training process can be skipped.

The advantage of DNNs over traditional methods is that they directly take the RGB images as inputs, and learn to find important regions and features by fitting the training data. In this way, manual feature engineering is no longer required, which enhances the scalability of such approaches. However, the classification tasks do not force models to encode rich product information commonly found in e-commerce applications. The resulting models can achieve good classification performance by only extracting information relevant to identifying pre-defined labels. One may propose to solve this issue by using finer-grained labels (e.g., defining each unique product as a class). However, this leads to a huge classification layer, which can incur high memory and computation costs during training.

\textbf{Deep Metric Learning-based Methods:} 
Compared to classification-based DNNs, deep metric learning (DML) can better preserve the rich product information in VMs when trained with large-scale fine-grained labels. This is because DML trains a CNN that directly outputs an embedding. The goal of DML is to make the data points within a class to be close together, and those from different classes far apart. The closeness measures can be defined in different ways. Since DML directly optimizes the data representations, the classification layer is not required in the training process. This reduces memory and computation costs, making training on large-scale fine-grained labels viable. The DNNs used for DML are the same as those under the classification-based deep learning methods, with the classification layer replaced by a fully connected layer to output the VM representations with a desired number of dimensions. To train such a model, a variety of loss functions have been proposed.

\emph{Proxy-based DML}: 
Proxy-based loss, which represents a group of data with a proxy, has been proposed for DML. The proxy is usually a trainable 1-D vector, which is trained together with the model through backpropagation. The loss can be calculated based on the similarities between the proxy and data points. For example, ProxyNCA loss \cite{movshovitz2017no} is defined as the logarithm of the Softmax function. The logits of the Softmax are the euclidean distance between the features of data $x_i$ and the trainable proxies of $x_i$.
In \cite{movshovitz2017no}, two proxy assignment strategies have been proposed: 1) static proxy assignment, under which each proxy is defined to belong to a fixed class (i.e., the relationships between proxies and classes do not change); and 2) dynamic proxy assignment, which assigns each data point to its closest proxy. The advantage of dynamic proxy assignment is that it does not require ground truth labels in order to train the model. On the other hand, static proxy assignment achieves better accuracy under supervised learning settings.


Proxy-based DML usually assigns proxies to classes via one-to-one mapping \cite{kim2020proxy,deng2019arcface,wang2018cosface,liu2016large,liu2017sphereface}. Nevertheless, approaches which assign multiple proxies for each class have emerged \cite{movshovitz2017no,qian2019softtriple}. \citeauthor{movshovitz2017no} propose to randomly assigns multiple classes to a proxy so that the number of proxies can be smaller than the number of classes \cite{movshovitz2017no}. This helps to tackle the high costs incurred by having a large number of classes, albeit at the cost of some performance degradation.


\emph{Pair-based DML}:
Pair-based loss is computed based on the similarities between pairs of data points. Two types of data pairs are usually considered in the loss: 1) positive pairs, in which the two data share the same ground truth labels; 2) negative loss, in which the two labels are different. The data pairs are usually formed within minibatch data. The most straightforward pair-based loss is the contrastive loss \cite{chopra2005learning}. It is the sum of the euclidean distance of positive pairs, minus the sum of distance of selected negative pairs. The selection of negative pairs is based on their distance. If the distance is smaller than $\lambda$, then it will be selected. $\lambda$ is a predefined hyperparamter. It helps the loss function to focus on the hard negative examples which appear close together. By minimizing the contrastive loss, the similarities between negative data pairs are minimized, and those between positive pairs are maximized.
Another commonly used pair-based loss is the triplet loss \cite{schroff2015facenet}. Instead of optimizing the similarity between each pair of data points, the triplet loss considers the relative difference between the distance among positive pairs and that among negative pairs.

Existing research works such as \cite{wang2019multi,sun2020circle,sohn2016improved,yi2014deep} attempt to assign high learning weights to hard examples (i.e., negative data pairs that have high similarity values, or positive data pairs that have low similarity values) so that the model can learn to represent hard examples better. \citeauthor{wang2019multi} \cite{wang2019multi} found that most pair-based models weight the negative data pairs based on either self similarities, positive relative similarities or negative relative similarities compared with other pairs. Specifically, the three types of similarities of negative data pairs are defined as follows:
\begin{itemize}
    \item \textbf{Self similarity}, which is the cosine similarity between the negative sample and the anchor. This is used in contrastive loss \cite{chopra2005learning}, binomial deviance loss \cite{yi2014deep} and MS loss \cite{wang2019multi}.
    \item \textbf{Positive relative similarity}, which is defined as the relative difference between the similarity of a negative data pair and that of positive data pairs. This is used in triplet loss \cite{schroff2015facenet}, histogram loss \cite{ustinova2016learning}, NCA loss \cite{salakhutdinov2007learning}, MS loss \cite{wang2019multi} and circle loss \cite{sun2020circle}.
    \item \textbf{Negative relative similarity}, which represents the relative difference between the similarity of a negative data pair and that of other negative data pairs. Examples include N-pairs loss \cite{sohn2016improved}, lifted structure loss \cite{oh2016deep}, NCA loss \cite{salakhutdinov2007learning} and MS loss \cite{wang2019multi}.
\end{itemize}
The similarities of positive data pairs can be defined in a similar fashion.

Pair-based loss functions enable training VM representation models without a classification layer. However, the accuracy heavily depends on training batch size, because the pairs are built from data within a minibatch. More data pairs are available if the batch size is large, which can better reflect the underlying data distribution. To reduce the dependency on the batch size settings, \citeauthor{wang2020cross} \cite{wang2020cross} proposed a memory bank which saves the data features obtained from previous iterations in a first-in-first-out queue. In each iteration, the features of current minibatch data is enqueued into the memory bank, while the oldest features are dequeued if the number of features reaches its maximum limit. By comparing the minibatch data with the memory bank data, the number of compared data pairs can be significantly increased. Thus, applying the memory bank with pair-based loss improves the accuracy of DML. 

\subsubsection{Feature Aggregation}
Feature aggregation uses multiple features of the local patches or frames in VMs to generate a compact VM representation. Typically, the output representation has a fixed dimension, while the number of input features is not fixed. Feature aggregation can be leveraged to obtain a global feature for an image (when inputs are local patch features) or a video (when inputs are frame features).

Bag of Visual Words (BoVW) \cite{csurka2004visual,sivic2003video} performs K-means on the features of local patches on all images in a dataset. Each local feature is then denoted by the ID of the cluster it belongs to, which is generated by K-means. Finally, BoVW represents each image as the histogram of local feature IDs. Vector of Local Aggregated Descriptors (VLAD) \cite{jegou2010aggregating} also applies K-means in the same way as BoVW. The sum of the residuals between the local features and its cluster centre is used as the feature of image $i$ belonging to the cluster $k$ (denoted as $V(i,k)$). Finally, $\left[ V(i,1), V(i,2),...,V(i,K) \right]$ is used as the global representation of the $i$-th VM.

To further improve VLAD,  NetVLAD \cite{arandjelovic2016netvlad} has been proposed to express $V(i,k)$ in a differentiable form so that it can be incorporated into DNNs. The clustering process is also skipped by replacing the cluster centre $c_k$ with a learnable parameter. NeXtVLAD \cite{lin2018nextvlad} has been proposed to further reduce the number of parameters compared to VLAD while maintaining model accuracy.

\subsection{Cognitive Property Assessment} \label{sec:cpa}
Cognitive properties \cite{goetschalckx2019ganalyze} (e.g., memorability, aesthetics, perceived emotion) are crucial for AIPVG. They represent a high-level understanding of the visual materials which is required for storyline generation. Accurately measuring the cognitive properties of VMs helps identify memorable and attractive VMs, which in turn, increase the persuasiveness of the generated videos \cite{okeefePersuasionTheoryResearch2016}.
Cognitive properties have been studied in-depth in fields such as psychology, sociology, and cognition science \cite{rolls2005emotion,weitz1956role}. There has been no mathematical formulation proposed for them, which makes it challenging to estimate the cognitive properties of images and videos using algorithms with hand-crafted features. In recent years, researchers start to build deep learning models to address this problem \cite{lee2019image}. 

\subsubsection{CNN-based CPA}
The baseline models to assess cognitive image properties are CNNs (e.g., AlexNet \cite{krizhevsky2014}, ResNet \cite{he_deep_2016} and VGG \cite{simonyan_very_2015}) which have been pre-trained on ImageNet \cite{deng_imagenet:_2009} and fine-tuned on CPA datasets. A CNN takes an image as the input, and outputs a score that measures each type of cognitive property. Existing research works \cite{xu2014visual,jang2021analysis} show that features learned from image classification tasks benefit cognitive image property assessment. To improve the traditional CNN architecture for CPA tasks, a variety of model components have been proposed.

\textbf{Fusing Features of Different Layers:} It has been found that combining features of different layers in a CNN can improve emotion classification performance \cite{zhu2017dependency,rao2020learning,yang2018retrieving}. This is because different layers refer to different parts of an image \cite{zhu2017dependency}. To fuse the features of different layers, a layer fusion component is needed in the existing CNN architectures. \citeauthor{zhu2017dependency} \cite{zhu2017dependency} designed this component as a bi-directional GRU, which is a variant of RNN \cite{cho2014learning}. Other works such as \cite{rao2020learning,yang2018retrieving} designed it as a sub-network consisting of convolutions and fully connected layers. These components can either be trained separately \cite{zhu2017dependency} or together with the CNN backbones \cite{rao2020learning,yang2018retrieving}.

\textbf{Avoiding Re-scaling:} The resolution and aspect ratio of an image are important when assessing its quality and aesthetics. Thus, it would be desirable if these two pieces of information can be preserved in CPA tasks. There are two main approaches for achieving this goal: 1) using pooling layers, or 2) using feature extractors.

Existing works such as \cite{ko_pac-net:_2018,murray2017deep,chen2020adaptive} leverage spatial pyramid pooling (SPP) \cite{he2015spatial} to avoid altering the resolution and aspect ratio of an input image. The SPP layer is able to pool arbitrarily-sized features $x$ and generate fixed-length outputs. It can be viewed as a layer composed of 3 adaptive pooling operations. Each adaptive operation takes $x$ as the input and outputs a fixed-sized matrix. Here, ``adaptive'' means that the kernel size $(w_{kernel},h_{kernel})$ and stride $(s_{width},s_{height})$ of the pooling depends on the input size and desired output size (i.e., $w_{kernel}=\lceil \frac{w_{input}}{w_{output}} \rceil$ where $w_{input}$ and $w_{output}$ are the widths of the input and the output, respectively). Similarly, the height of the kernel can be defined. Finally the 3 pooled features are flattened and concatenated to become the output of the SPP layer.

Besides, \citeauthor{chen2020adaptive} \cite{chen2020adaptive} proposed a component that linearly combines features extracted via two dilated convolution operations at different dilation rates so as to preserve information related to aspect ratios. The feature combination can be viewed as a variant of (weighted) average pooling, where the weights are calculated based on the aspect ratio and dilation rate. \citeauthor{ma_-lamp:_2017} \cite{ma_-lamp:_2017} proposed A-Lamp, which selects multiple patches using a patch selection sub-module, VGG16 is used to extract features from the unscaled selected patches. The resulting features are the concatenation of max-pooled and average-pooled features of multiple patches.

Another line of research first extracts features by using feature extractors, and then uses a regressor to predict the final score. The extractor can either be expert-designed \cite{tu2021efficient,tu2021rapique} or pretrained neural networks \cite{ying2021patch,hosu2019effective}. The extractors are not fine-tuned on the target dataset to reduce time and memory costs. After obtaining the features, a regressor which takes the features as the inputs and outputs the score is trained. \citeauthor{ying2021patch} \cite{ying2021patch} proposed to use 2D frame-level features and 3D video-level temporal features to extract video features. The 2D features are extracted by a image-quality-aware CNN called PaQ-2-PiQ \cite{ying2020patches}, while the 3D features are extracted with a 3D ResNet-18 \cite{hara2017learning} pre-trained on the Kinetics dataset. InceptionTime \cite{fawaz2020inceptiontime} is utilized as the regressor network, which is a CNN that is designed for time series regression. A follow-up work \cite{tu2021ugc} uses the features extracted by multiple video quality assessment networks that are publicly available, and then uses the SVM as the regressor.

Both approaches successfully avoid re-scaling in CPA tasks, albeit with reduced model throughput or accuracy. For most pooling-based techniques, models without re-scaling can only be trained on a batch size of 1, thereby consuming a huge amount of time. For models that first extract features with a pretrained model, the training is not in an end-to-end fashion. The feature extractor is not fine-tuned on the target dataset, which leads to accuracy drops.

\textbf{Paying Attention to Local Regions:} Research on emotion recognition and aesthetics score estimation has found that mining the local regions improves the model performance. A line of works explicitly identify regions of interest (ROIs) and estimate the cognitive properties based on the ROI features. \citeauthor{ma_-lamp:_2017} \cite{ma_-lamp:_2017} designed a region proposal method with the consideration of professional photography rules and human visual principles. The method estimates the interests based on saliency, pattern diversity and spatial distance among selected regions. \citeauthor{yang2018visual} \cite{yang2018visual} and \citeauthor{rao2019multi} \cite{rao2019multi} use existing region proposal methods (e.g., EdgeBoxes \cite{zitnick2014edge} and Faster R-CNN \cite{ren2015faster}) that take an image as the input and produce multiple bounding boxes that describe ROIs. However, they are not trained on target datasets, which might not always achieve acceptable performance. Alternatively, an input image can be directly split into grids and take each grid cell as a region \cite{liu_composition-aware_2019,she2021hierarchical}.

To obtain the final predictions from the regions, \cite{liu_composition-aware_2019,she2021hierarchical} proposed to use graph convolutional operations to reason on the graph of ROIs. Each node in the graph represents an ROI, while each edge represents the similarity between two ROIs. \citeauthor{yang2018visual} \cite{yang2018visual} proposed to linearly combine global and local predictions as the final results. \citeauthor{rao2019multi} \cite{rao2019multi} proposed to concatenate the ROIs and global features and send them to a classifier to predict the results. \citeauthor{wei2021user} \cite{wei2021user} first performs prediction on each local region, then aggregates the predictions by combining both the results of score fusion and Top-K decision fusion.

Another line of research employs the attention mechanism \cite{vaswani2017attention} to implicitly mine local regions. The attention mechanism allows the inputs to interact with each other, and allocates its attention to the inputs \cite{vaswani2017attention}. The outputs are aggregates of these interactions and attention scores. The attention mechanism has been used in existing CPA tasks to identify and highlight important local regions of a given VM. Depending on the dimensions involved, the attention mechanisms for CPA models can be divided into spatial attention \cite{yang2018weakly,zhao_pdanet:_2019,cohendet2019videomem,sheng_attention-based_2018} and channel-wise attention \cite{zhao_pdanet:_2019}. The spatial attention calculates an importance score for each vector in the spatial dimension, while the channel-wise attention generates a score for each channel. 

\subsubsection{CPA Model Optimization}
Apart from the model architecture, model optimization methods are also crucial to CPA. The conventional optimization approach is to minimize the mean squared error (MSE) loss \cite{ding2019social,xu2020multimodal,kong2016photo,shu2020learning} for regression tasks. To improve the MSE loss, \citeauthor{murray2017deep} \cite{murray2017deep} proposed the Huber loss to improve model robustness to outliers. \citeauthor{zhao_pdanet:_2019} \cite{zhao_pdanet:_2019} proposed the polarity-consistent regression (PCR) loss for emotion regression. It assigns a penalty to the predictions of a sample that has opposite polarity to the ground truth. The penalty is applied to the loss value of each sample in the form of MSE. The cross-entropy loss is commonly adopted by works that treat CPA as a classification task \cite{ma_-lamp:_2017,sheng_attention-based_2018,liu_composition-aware_2019}. \citeauthor{talebi2018} \cite{talebi2018} proposed the Earth mover's distance (EMD) loss to improve classification performance. Compared to the cross-entropy loss, EMD accounts for the distance between categories. EMD is defined as the minimum cost to move the mass of one distribution to another \cite{talebi2018}. In addition, instead of improving the MSE or the cross-entropy, an alternative is to use pairwise-based loss (e.g. triplet loss, ranking loss) to optimize the model \cite{schwarz_will_2018,kong2016photo,ding_intrinsic_2019}.


Moreover, multi-task learning is adopted by \cite{shu2020learning,kao2017deep,kong2016photo} in which auxiliary training tasks are incorporated to provide extra supervision to the models, thereby enhancing CPA performance. Models reported in \cite{shu2020learning,kao2017deep,kong2016photo} are trained on the AADB dataset that provides scores of aesthetics-related attributes (e.g., colour harmony, vivid colour, good lighting) \cite{shu2020learning,kong2016photo}. The model jointly outputs the prediction of the aesthetics score as well as the attribute score. 

\subsection{Discussion}
In this section, we introduced VM understanding methods for visual storytelling, including VM representation and cognitive property assessment (CPA). On one hand, well-learnt representations can enhance the central route of persuasion \cite{petty_elaboration_1986} by improving the informativeness of storytelling. On the other hand, an accurate CPA model can improve the peripheral route of persuasion \cite{petty_elaboration_1986}. In the next section, we further elaborate how these considerations make an impact on visual storyline generation.

\section{Visual Storyline Generation}






Visual Storyline Generation (VSG) approaches take the information of the VMs as inputs in order to optimize the selection and sequencing of the VMs to produce visual storylines. 
Based on the criteria adopted for VM selection and sequencing, we divided the existing VSG methods into three categories: 1) Spatial-Temporal VSG, which use the time and location information as criteria to perform VSG; 2) Data-Driven VSG, which attempt to train a VM selection or sequencing model based on data; and 3) Knowledge-Based VSG, which perform selection and sequencing of VMs based on relevant domain knowledge (e.g., plot analysis of written stories or film making principles).

\subsection{Spatial-Temporal VSG}
The timestamp information and location information related to VMs have been leveraged to provide intuitive storytelling structures by existing VSG approaches \cite{hua_photo2videosystem_2006,wu2016monet,chen_tiling_2006}. 
Spatial-Temporal VSG approaches are well-suited to photos which have been taken by users while traveling. These approaches commonly involve two major steps: VM selection and VM sequencing.

\subsubsection{VM Selection} 
The aim of the VM Selection step is to identify and discard low quality and duplicate VMs. Earlier works define low quality images as underexposed, overexposed or blurry photos \cite{chen_tiling_2006}. \citeauthor{hua_photo2videosystem_2006} \cite{hua_photo2videosystem_2006} further considers the homogeneity among candidate VMs for selection. Based on \cite{hua_photo2videosystem_2006}, \citeauthor{wu2016monet} \cite{wu2016monet} further added sharpness and features of important patches as selection criteria.
All three lines of works use a wavelet-based blur detection method reported in \cite{tong2004blur} to measure the degree of blur in VMs. In \cite{hua_photo2videosystem_2006}, underexposure, overexposure and homogeneity are measured by color entropy (i.e., the Shannon entropy of normalized HSV color histogram). Nevertheless, these approaches can be computationally costly. Thus, more straightforward methods to measure the underexposure and overexposure by counting the number of dark and bright pixels in a photo have been proposed \cite{chen_tiling_2006,wu2016monet}. If the intensity value of a pixel is higher than a designated threshold, it is considered as bright; if the value is lower than a predefined threshold, it is considered as dark.

Existing works have leveraged the above-mentioned evaluation metrics to design policies for filtering VMs. In \cite{chen_tiling_2006}, a threshold is set for each of these metrics. If any photo scores below any threshold, it is discarded. \citeauthor{hua_photo2videosystem_2006} \cite{hua_photo2videosystem_2006} adds two more metrics (i.e., color entropy and the degree of blur) into consideration. Photos with final scores below a predefined threshold are considered as low quality and removed. \citeauthor{wu2016monet} \cite{wu2016monet} treats these metrics as features, and uses an SVM to determine VM quality. The SVM is trained on a large image set with each image manually labeled as ``good'' or ``bad'' quality.
In addition, \cite{hua_photo2videosystem_2006,chen_tiling_2006} use the temporal information to group the VMs. Then, for each time segment, a content-based clustering algorithm is used to generate clusters of VMs. The top $K$ clusters are selected as the candidate VMs for generating the storyline. The importance of each cluster is also calculated by jointly considering the number of photos in the cluster and the within-cluster distance. 

Another important consideration during VM selection is duplication filtering. For temporally adjacent photos, \cite{hua_photo2videosystem_2006,chen_tiling_2006} calculate the pairwise euclidean distance on the down-sampled images to determine if they are duplicates. \citeauthor{wu2016monet} \cite{wu2016monet} uses a feature extraction method \cite{winder2007learning} to obtain features and calculates pairwise similarity. If a pair has a small distance, the image with a lower quality score will be removed \cite{hua_photo2videosystem_2006,chen_tiling_2006,wu2016monet}. \citeauthor{wang2012generating} \cite{wang2012generating} proposed a graph-based solution to select representative images for building a visual storyline. A graph is first built in which each node represents an image with a caption, and each edge represents the similarity between a pair of images considering both visual and textual components. The VM selection task is then performed by minimum-weight dominating set finding in the graph.

\subsubsection{VM Sequencing} 
Spatial-Temporal VSG approaches generally sequence the selected images according to their timestamps, while placing photos that are taken in close spatial proximity together \cite{hua_photo2videosystem_2006,chen_tiling_2006,wang2012generating}.
\citeauthor{chen_tiling_2006} \cite{chen_tiling_2006} presents the photos in each cluster as a tiling slideshow. Each frame is a tiling slide defined by a template that contains several cells, with each cell containing one photo. The method dynamically determines the template for each cluster. The photo importance (PI) score and template importance (TI) score have been proposed to provide more fine-grained control over the sequencing step. PI is calculated by considering the face region ratio and an attention value \cite{ma2002user} near the center of an image. TI is defined as the ratio between the area of a cell over the entire area. The PI values for all photos in a cluster are sorted in descending order, and thus becomes a cluster feature. The template feature is obtained in a similar fashion. The template is eventually determined by finding the one that maximizes the cosine similarity between the cluster feature and template feature. After the template is selected, photos are placed in cells according to their importance scores (i.e., a photo that ranks $i$-th by its importance score is placed into the $i$-th cell).

\subsection{Data-Driven VSG}

Another line of research performs visual storyline generation using machine learning models trained on large-scale datasets.

\subsubsection{Unary Models}
\citeauthor{agrawal2016} \cite{agrawal2016} proposed a solution to sort image-caption pairs into storylines. A unary model has been proposed, which takes a permutation of $n$ image-caption pairs as the input and outputs the appropriateness of the permutation. Specifically, a VGG19 model \cite{simonyan_very_2015} is used to embed images and a GRU \cite{cho2014learning} trained on the BookCorpus dataset \cite{kiros2015skip} is used to embed the captions. Both embeddings are concatenated and fed into an MLP. During training, only the parameters of the MLP layers are updated. The best permutation is found with the Hungarian algorithm \cite{munkres1957algorithms}.

In contrast, \cite{sigurdsson2016} proposed a variant of RNN, namely skip RNN, to construct a visual storyline in a bottom up manner by iteratively predicting the next image to be placed in a storyline. To perform image selection, the skip RNN model takes a series of photos as the input. At step $t$, the features of the currently selected images are taken as the input, and the model outputs a feature $y$. Softmax operation is performed on the dot product of $y$ and features of subsequent images $[x_{t+1},x_{t+2},...,x_n]$ to calculate the probability of image being $x_j, j \in [t+1,n]$ to be the next image. In addition, the skip RNN can also be used in the sequencing step by predicting the next image that is not selected in the output sequence yet.

\subsubsection{Pairwise Models}
Pairwise comparison-based models have been proposed to perform the VM sequencing steps. They generally take a pair of elements $(i,j)$ as the input and predict whether $i$ should be placed before $j$.

Existing works such as \cite{xu_self-supervised_2019,siar2020unsupervised,zhukov2020learning,hu2021contrast} aim to learn a good visual representation for VM sequencing in an unsupervised fashion. The authors utilize a large-scale video dataset and divide each video into several clips. The task is to sequence the video clips by inferring their temporal dependency. \citeauthor{xu_self-supervised_2019} \cite{xu_self-supervised_2019} proposed a solution that first uniformly samples clips from the video. Then, a 3D CNN is used to extract features of each clip. Finally, the ordering of the video clips is established by an MLP layer. The MLP layer takes the concatenation of embeddings of the video clip pairs as the input, and outputs features for each possible pair. The features are further concatenated as the final representation, to produce the probability distribution over all possible sequences. The training loss is the cross-entropy, in which each possible permutation is considered as a class.

In \cite{xu_self-supervised_2019}, the number of clips is a predefined hyperparameter. The approach can only support between 2 to 5 video clips, because a large number of clips leads to a huge number of permutations, thus making the classification very difficult. \citeauthor{siar2020unsupervised} \cite{siar2020unsupervised} proposed to solve this issue by grouping the frames. A total of $\frac{N!}{2}$ groups can be formed. During training, the classification layer is only required to choose one of the two possible permutations.

\citeauthor{el2019skip} \cite{el2019skip} noticed that for actions like doing pull-ups, the reverse order of the pulling up frames in a video is also temporally plausible. To address the problem of label ambiguity caused by this problem, they proposed a solution called skip-clip to take contextual information into consideration to improve order prediction. The context is defined as a clip that is $r$ frames before the first target clip. During training, instead of using classification, a hinge rank loss is used, which aims to ensure that the target clips that are temporally close to the context clip have high cosine similarity values.



\subsection{Knowledge-Based VSG}
A major limitation of data-driven VSG approaches is the heavy reliance on relevant data, which might not be readily available. Therefore, most existing data-driven VSG works leverage temporal visual storylines as the training data, which can be obtained relatively easily by cutting up existing videos into clips. For other VSG tasks that consider more than temporal relationships (e.g., visual coherency, persuasiveness, viewing experience), there is no publicly available dataset to support data-driven VSG approaches. Collecting such a dataset is difficult and expensive because the criteria involved tend to be subjective.

To address this issue, some works incorporate knowledge from other domains to perform visual storyline generation. In this section, we describe the goals, domain knowledge selected, objective function definitions based on such knowledge, and the optimization approaches of these knowledge-based VSG approaches.

\subsubsection{Domain Knowledge from Storytelling Research}
Storytelling plot analysis has been incorporated into personal video clip sequencing \cite{choi_video-story_2016,zhong2018}. They propose that visual storylines should follow a common plot like other written stories. Such a plot template starts with exposition, then moves on to rising action followed by a climax, and finally reaches a resolution. The authors define this plot using the metric of dynamicity. At the beginning of storylines, the content dynamicity is relatively low. Then, it gradually increases until the climax. Finally, it decreases but still remains higher than the dynamicity at the beginning. The dynamicity is calculated as the average magnitude of the dense optical flow features \cite{liu2008human}. In the context of persuasion, the dynamicity is useful when applying the peripheral route persuasion under the Elaborate Likelihood Model (ELM) \cite{petty_elaboration_1986}.

In addition, \citeauthor{choi_video-story_2016} \cite{choi_video-story_2016} proposed to consider coherence during the sequencing step. The dissimilarity between two adjacent video clips is utilized to calculate coherence for a possible sequence. The final objective function is the linear combination of coherence and dynamicity. \citeauthor{zhong2018} \cite{zhong2018} calculates coherence using two RNNs. The RNNs are trained in an unsupervised manner, which aims to select the next video clip from the set of clips that are not yet selected. The video clips are obtained in a similar way as the temporal-based data-driven VSG \cite{xu_self-supervised_2019}. One of the RNNs takes the spatial pyramid pooling on the histogram of dense optical flow (SPP-HOOF) as the input. The other RNN takes the C3D feature of current video clips as the input. They each output a probability distribution for the next clip. Then, the two distributions are added together to form the final coherence vector.

\subsubsection{Domain Knowledge from Persuasive Theory} \citeauthor{liu_generating_2019} \cite{liu_generating_2019} attempts to generate visual storylines that can maximize persuasiveness. According to the research on persuasive advertising \cite{armstrongPersuasiveAdvertisingEvidencebased2010}, informativeness, attractiveness and emotion are key stimuli for achieving persuasion. In the ELM persuasion model \cite{petty_elaboration_1986}, informativeness measures whether the provided information is sufficient for the central route of persuasion, while attractiveness and emotion are important for the peripheral route of persuasion. The authors propose to use dissimilarities between adjacent visual materials, aesthetics \cite{talebi2018} and arousal scores to compute informativeness, attractiveness and emotion, respectively. Both aesthetics and arousal scores have been described in Section~\ref{sec:cpa}. A Learnable Wundt Curve (LWC), which is inspired by the Wundt Curve \cite{berlyne1960}, has been proposed to fuse the scores and produce a final persuasive score. The Wundt Curve is a bell-shaped curve in which increases in persuasion-related stimuli first cause the persuasiveness of the videos to increase. Once these stimuli increase past certain values, further increasing them causes the persuasiveness of the videos to decrease rapidly. Such a theoretical curve indicates the effect of over stimulation on viewers' experience. \citeauthor{liu_generating_2019} \cite{liu_generating_2019} uses two transformed Sigmoid functions to translate the Wundt Curve into a machine learning model, in which the key parameters are learned via backpropagation. The LWC is also used in \cite{liu2020generating} to calculate persuasion scores.

\subsubsection{Domain Knowledge from Film-making Principles} \citeauthor{liu2020generating} \cite{liu2020generating} incorporates film-making principles with the aim of enhancing the viewing experience of the generated videos. This can, in turn, improve the persuasiveness of the videos under the peripheral route of persuasion. Three principles are considered: 1) the visual storyline should start with a wide shot and gradually narrow to a close-up shot; 2) the visual storyline should have a logic, and thus be easy for viewers to follow; and 3) there should be graphic discontinuities between scenes to keep viewers excited. To implement these principles algorithmically, the VMs are first clustered. Each cluster is treated as a shot in the video. In this way, a preliminary logic flow for the visual storyline can be created. The objective function for the sequencing step is a linear combination of distance penalty and graphic discontinuity. The distance penalty considers two distance functions: 1) the Salient Region Ratio (SRR), which measures the distance in terms of attention, which is calculated by the attention map generated by the CNNs and SIFT; and 2) the semantic distance (SED), which measures the distance between the shots and the product which is being promoted in the video. It is measured by the text embedding distance between the predicted classes and the text description of the product. The distance penalty function produces a penalty when the SRR decreases and the SED increases over time. By minimizing the penalty, Principle 1 can be achieved. The graphic discontinuity is represented by the cosine dissimilarities between features of adjacent shots. 


Finding a sequence that minimizes the objective function in VSG is an NP-hard problem. In \cite{choi_video-story_2016}, a branch-and-bound searching approach is proposed, which reduces the search space by updating the lower and upper bound scores for each subspace, and removes all subdivided sub-spaces that the lower bound is larger than the upper bound. \citeauthor{zhong2018} \cite{zhong2018} proposes a sub-modular ranking method, which iteratively selects the next video clip that shares high coherence with other clips while increasing the dynamics. \citeauthor{liu_generating_2019} \cite{liu_generating_2019} introduces a dynamic programming method, which formulates the searching process as a group backpacking problem. Each group is considered as a cluster of VMs, and the items in a group are a sub-sequence with a specific length. The goal is to maximize the persuasiveness within a predefined video length limit.

\subsection{Discussion}
Visual Storyline Generation selects and sequences the visual materials to form the storyline of the video, which is the most important step in AIPVG. In the context of E-commerce product promotion, it is more reasonable to use visual content instead of the time and location information in the images or video clips to form the storylines, as the main purpose of the generated the videos is to persuade viewers. In the context of E-commerce, the sellers always edit, crop, and concatenate photos before putting them online, and E-commerce platforms delete the EXIF data to minimize the size of files. These factors made the date and location information of the VMs unusable and render the Spatial-Temporal VSG approaches ineffective. Data-driven VSG can be a promising direction. However, current data-driven works focus more on inferring temporal relations between video clips. No dataset about persuasive storylines is publicly available, which hinders the research and deployment of data-driven VSG for persuasion. The knowledge-based VSG approaches can be more readily deployed in real-world systems. However, the performance of such approaches may not be as good as the data-driven VSG methods. In addition, they are not well suited for personalizing storylines due to limited learning capability.

\section{Post-Production}
The post-production phase often follows the visual storyline generation phase with the aim to further enhance the overall viewing experience of the resulting videos. In this phase, edits on the selected visual materials, including generating visual effects and background music, are often performed. Finally, the videos are rendered using software such as MoviePy\footnote{\url{zulko.github.io/moviepy}} and VidGear\footnote{\url{abhitronix.github.io/vidgear}}. 

\subsection{Music Recommendation \& Generation}
Music recommendation approaches aim to recommend background musics from existing music repositories based on video contents. This is generally performed as a feature extraction and matching task between videos and musics. 
The approach in \cite{kuo2013background} first extracts features from videos and musics (e.g., Saturation Proportion and Contrast for videos, Beat Histogram and Loudness for musics). Then, it performs K-means to discretize the features and mine the correlations between discretized video and music features using a modified version of Multiple-type Latent Semantic Analysis (M-LSA) \cite{wang2006latent}. The recommendations are generated by finding the nearest neighbours of the projected video features from the database of projected music features. 
\citeauthor{liu2018background} \cite{liu2018background} proposed to extract low-level features for both videos and musics, paying special attention to the emotional undertones of the video contents and the musics. The relevance scores are then calculated for recommendation, based on a custom designed emotion-aware scoring function.
In \cite{lin2016automatic,lin2017automatic}, visual features are extracted by averaging the frame features produced by a pretrained VGG16 model. A multi-task learning method is proposed to train a network that jointly generates the music features and analyzes the emotions and styles of the musics. Music retrieval is then performed by a regression model, which computes the similarities between the generated music features and the musics in the repository. 
The Cross-modal Variational Auto-encoder (CMVAE) \cite{yi2021cross} constrains the latent variables of the musics and videos while jointly generating these latent variables. An embedding for each video or music is generated by the encoder of CMVAE. Finally, the model recommends musics by calculating the similarity scores between the music and video embeddings.

Recommending musics from repositories risks infringing music copyrights. Thus, another line of research attempts to address this issue by training DNNs to generate musics based on given video contents.
In \cite{gan2020foley,su2020audeo,su2020multi}, approaches have been proposed to generate musics for silent videos containing a musician playing a musical instrument. However, the application scenarios of such techniques are not well aligned with promotion video generation in e-commerce. 
\citeauthor{di2021video} \cite{di2021video} proposed to generate background music based on given video contents. During training, a transformer is used to analyze the attributes of Musical Instrument Digital Interface (MIDI) musics. The attributes are defined by compound words. The transformer decoder uses the sequence of words, along with beat timing information, to learning to re-generate the input MIDI musics. During inference, the attributes of input videos (e.g., motion speed, motion saliency and timing) are also denoted by the compound words heuristically. Then, the trained decoder outputs the music.

\subsection{Still Image Animation}
Still image animation aims to create a short video clip with realistic motions based on given image(s). \citeauthor{mathieu_deep_2016} \cite{mathieu_deep_2016} proposed to predict future images from a video sequence. A multi-scale architecture is designed, in which the images are fed into a network with size $k/2$. Then, the outputs are aggregated with the result produced by the network with size $k$. \citeauthor{ferrari_learning_2018} \cite{ferrari_learning_2018} extended this idea and proposes a two-stage generation framework in which videos are generated from structures and then refined by temporal signals. \citeauthor{shaham_singan:_2019} \cite{shaham_singan:_2019} proposed SinGAN, which performs single image animation using a pyramid of GANs. Each level takes the output of the previous level, as well as a random noise as the input to produce refined images with higher resolutions. 

Other works focus on animating images involving human bodies by adding visual effects. \citeauthor{chan_everybody_2019}  \cite{chan_everybody_2019} proposed a model that leverages a reference video that shows a person performing an action, and a photo with the target person. The model transfers the reference moves to the target person to generate animations. The model extracts the pose from the reference video as a sequence of skeleton images, then computes the pose-to-appearance mappings to generate the frames of the target person for each skeleton image. Temporal coherence is considered when training the generator. \citeauthor{yoon2021pose} \cite{yoon2021pose} further proposed to improve pose-based human image animation by reducing the visual artifact of the output videos. Sub-networks are trained to explicitly predict silhouette, garment labels and textures with synthetic data. 

\subsection{Discussion}
The post-production step improves the attractiveness of the AI generated promotional videos. Music generation and recommendation enhances the sense of immersion, while still image animation enhances the dynamism of the videos. Together, they serve to enhance the peripheral route of persuasion under ELM. 

However, excessive use of post-production techniques might negatively affect the persuasiveness of the generated videos under the central route of persuasion under ELM. Improper still image animation might result in incorrect information being presented to the viewers as the animation transformation is learned from the training data. For example, for a new product, some information might not be present in the input image, but is required to generate informative video clips. This makes still image animation ambiguous, reducing the informativeness of the video. An alternative solution that can enable the post-production step to enhance the central route of persuasion is to extract selling points \cite{Xu-et-al:2022IAAI} from the VMs or descriptions by summarization or image/video captioning techniques, then explicitly present this information as voice-over or captions in the generated videos. 

\section{Evaluation}
In this section, we discuss the techniques and methods for evaluating AIPVG approaches. Specifically, we introduce the evaluation metrics and datasets for representation learning and cognitive property assessment. We also summarize online testing methods for evaluating the generated storylines, the post-production approaches and the final generated videos.

\subsection{Evaluating VM Representations}
The quality of the VM representations can be assessed individually using a labelled test set. The evaluation is typically conducted in a retrieval fashion. In this section, we first introduce the publicly available datasets for VM representation learning, then discuss the evaluation metrics.

\subsubsection{Datasets}
\begin{table}[]
\small
\begin{tabular}{|l|l|l|l|}
\hline
Dataset                  & Instance & \# image/video            & \# class \\ \hline
Stanford Online Products \cite{oh2016deep} & Furniture & 59,551               & 11,318   \\ \hline
DeepFashion \cite{liu2016deepfashion}             & Clothing & 239,557              & 33,881   \\ \hline
DeepFashion2 \cite{ge2019deepfashion2}             & Clothing & 224,114              & 45,417   \\ \hline
Street2Shop \cite{hadi2015buy}              & Clothing & \textgreater 420,357 & 204,795  \\ \hline
MovingFashion \cite{godi2021movingfashion}  & Clothing & $\sim$ 15,045 &  15,045  \\ \hline
\end{tabular}
\caption{Publicly Available Products Datasets for Evaluating VM Representations}\label{tab:VMr}
\end{table}

The representation learning datasets for e-commerce product VMs mainly contain furniture and clothing. The Stanford Online Products dataset \cite{oh2016deep} contains images of furniture items on eBay. Others \cite{liu2016deepfashion,ge2019deepfashion2,hadi2015buy} provide clothing photos taken by consumers and sellers. Generally the photos from sellers are of better quality than those from consumers. Thus, the datasets can be used to assess the VM representations under different image qualities. Street2Shop dataset \cite{hadi2015buy} contains photos from ModCloth\footnote{\url{https://modcloth.com/}}, while DeepFashion \cite{liu2016deepfashion} contains images collected from Forever21 \footnote{\url{https://www.forever21.com}} and Mogujie \footnote{\url{https://www.mogujie.com}}. All of them are E-commerce platforms selling clothing. MovingFashion \cite{godi2021movingfashion} is a publicly available video dataset for clothing items. Each video is associated with a distinct image of a shop, and is retrieved from Net-A-Porter\footnote{\url{https://www.net-a-porter.com/}}, Instagram or Tik Tok.

\subsubsection{Metrics}
Features of each VM in the test set are extracted using the learned DNNs. Then, for each query VM in the test set, the $k$ nearest neighbours (NNs) are retrieved. The score is then calculated using the labels of $k$ NNs and that of the query VM. For instance, Recall@k is calculated as the ratio of query VMs such that at least one of their $k$ NNs has the same label as that of the query VM.

Recall@k is commonly adopted by VM representation research works as the evaluation metric \cite{wang2019multi,wang2020cross,qian2019softtriple,sun2020circle,liu2021noise,musgrave2020metric}. However, analysis has revealed the weaknesses of the Recall@k \cite{musgrave2020metric}. Improper embedding space designs can lead to artificially high recall@k scores. To address this issue, \citeauthor{musgrave2020metric} \cite{musgrave2020metric} has proposed R-Precision and Mean Average Precision at R (MAP@R) as alternative evaluation metrics. Denoting the total number of VMs belonging to the class of the query VM $y_q$ as $R$, the $R$ nearest neighbours can be retrieved. Suppose $r$ is the number of nearest neighbours that belongs to $y_q$. R-Precision is calculated as $\frac{r}{R}$. MAP@R considers the ranking of the correct retrievals. It is defined as mean average precision of the $R$ nearest neighbours for each query.

Another popular evaluation metric is Normalized Mutual Information (NMI) \cite{oh2016deep,oh2017deep,sohn2016improved}. NMI, which is defined as the ratio between the mutual information and the square root of the product of entropy values between the ground truth labels and clustering results, is originally designed to assess clustering quality \cite{schutze2008introduction}. As DML-based VM representation learning approaches carry out clustering on the representations extracted by the learned CNN models, NMI has been adopted to evaluate the performance of such approaches.

\subsection{Evaluating Cognitive Property Assessment Models}
The evaluation metrics of CPA models depends on the type of labels provided in the dataset. When the model is trained via classification, accuracy is commonly used as the main evaluation metric. While Spearman correlation coefficient (SPCC), mean absolute error (MAE) or mean squared error (MSE) are commonly adopted for evaluating the performance of regression-based CPA models. In this section, we focus on discussing the publicly available datasets for CPA evaluation.

There are a large number of publicly available datasets related to aesthetics, emotion and memorability which can be used for CPA evaluation. Here, we only list large-scale datasets containing more than 10,000 images (Table \ref{tb:2}) or more than 1,000 videos (Table \ref{tb:3}).

\begin{table}[]
\resizebox{1.0\linewidth}{!}{ 
\begin{tabular}{|l|l|l|l|l|l|l|}
\hline
             &                       & Task           & \# image & Source                            & \# category & Annotation                         \\\hline
Emotion      & The Flickr CC \cite{borth_large-scale_2013} & classification & 487,227        & Flickr                            & 24          & Extracted from image tag           \\\hline
Emotion             & \cite{yang_how_2014}  & classification & 354,192        & Flickr                            & 6           & Extracted from comments            \\\hline
Emotion             & FI \cite{you_building_2016}            & classification & 90,000         & Flicker and Instagram             & 8           & Crowdsourcing                      \\\hline
Emotion             & IESN \cite{zhao_predicting_2018}                 & regression     & 1,012,901      & Flickr                            & N.A.        & Extracted from text           \\\hline
Emotion             & LDL  \cite{yang_learning_2017}                 & classification & 20,745         & Flicker and Twitter               & 8           & Crowdsourcing                      \\\hline
Emotion             & CGnA \cite{kim_building_2018}                 & regression     & 10,766         &  Flickr                                 & N.A.        & Crowdsourcing                      \\\hline
Emotion             & LUCFER \cite{balouchian_lucfer:_2019}               & classification & 3,605,101      & Bing Search                       & 8           & Crowdsourcing                      \\\hline
Emotion             & WEBEmo \cite{panda2018contemplating}               & classification & 268,000        & Multiple Websites                 & 25          & Queries                            \\\hline
Emotion             & T4SA \cite{vadicamo2017cross}                 & classification & 1,473,394      & Twitter                           & 3           & Extracted from comments            \\\hline
Emotion             & MVSO  \cite{jou2015visual}                & classification & $\sim$7.36M    & Flickr                            & 24          & Queries + Crowdsourcing            \\\hline
Aesthetic   & AROD  \cite{schwarz_will_2018}                & regression     & $\sim$380K     & Flickr                            & N.A.        & Crowdsourcing                      \\\hline
Aesthetic             & AADB \cite{kong2016photo}                 & regression     & 10,000         & Flickr                            & N.A.        & Crowdsourcing                      \\\hline
Aesthetic             & AVA  \cite{murray2012ava}                 & regression     & 255,530        & DPChallenge.com                   & N.A.        & Multiple Experts rating            \\\hline
Aesthetic             & CUHK-PQ \cite{tang_content-based_2013}              & classification & 17,690         & professional photography websites &  2           & Crowdsourcing                      \\\hline
Aesthetic             & Flickr-AES \cite{ren2017personalized}              & regression & 40,000         & Flickr &  N.A.           & Crowdsourcing                      \\\hline
Aesthetic   & I2PA   \cite{ding_intrinsic_2019}               & regression     & $\sim$2.5M     &  Instagram                  & N.A.        & number of likes                    \\\hline
Memorability & LaMem \cite{khosla_understanding_2015}                & regression     & 60,000         & varouis public datasets            & N.A.        & Crowdsourcing\\\hline
\end{tabular}}
\caption{Image-based Cognitive Property Assessment Dataset}\label{tb:2}
\end{table}

\subsubsection{Emotions} Many large-scale datasets are built for emotion assessment that aims to detect the induced emotion given an image/video, i.e., what emotion will be aroused after the viewer sees the image. Based on how the dataset defines emotions, the emotion assessment can either be a classification task or regression tasks.

Many datasets \cite{borth_large-scale_2013,yang_how_2014,you_building_2016,yang_learning_2017,panda2018contemplating,vadicamo2017cross,jou2015visual} contain emotion labels for each image/video. The emotion category system varies across datasets (e.g., Plutchik's Wheel of Emotions \cite{plutchik1980general} used by \cite{borth_large-scale_2013,jou2015visual}, Ekman's six emotions \cite{ekman1971constants} in \cite{yang_how_2014,xu2016heterogeneous}, eight emotions defined by \cite{mikels2005emotional} are used for \cite{you_building_2016,yang_learning_2017,balouchian_lucfer:_2019,jiang2014predicting}, and Parrott's wheel of emotions \cite{parrott2001emotions} in \cite{panda2018contemplating}). In terms of the annotation method, the majority uses crowdsourcing to label the images \cite{you_building_2016,yang_learning_2017,balouchian_lucfer:_2019}. Other datasets are built by extracting emotions from the comments/text/tags for the images \cite{borth_large-scale_2013,yang_how_2014,zhao_predicting_2018,vadicamo2017cross}. There are also works which skip annotation by directly using the emotion categories as keywords when querying the image search engine \cite{panda2018contemplating,jou2015visual}. In addition, \citeauthor{panda2018contemplating} \cite{panda2018contemplating} found that existing datasets contain significant biases. Thus, they have collected a large-scale web image dataset to minimize the effect of dataset bias. \citeauthor{jou2015visual} \cite{jou2015visual} attempts to understand the relationships between emotions and languages. Twelve languages are considered when collecting images on Twitter so that culture-specific and inherent linguistic contexts can be taken into account by the dataset.

The above-mentioned datasets are well-suited for image-based emotion classification. 
However, to be useful for visual storyline generation, the distinct class labels require further processing (e.g., embedding).
Some datasets assign VAD scores for each image \cite{yang_learning_2017,kim_building_2018,zhao_predicting_2018}. \citeauthor{zhao_predicting_2018} \cite{zhao_predicting_2018} extracts these scores from the keywords in the texts related to the images (i.e., title, tags and descriptions) using the VAD values of 13,915 English lemma \cite{warriner2013norms}. Extracting VAD values from text is a way to build large-scale datasets in an affordable manner. However, such an approach often sacrifice accuracy to some extent since user-generated texts might not always reflect their real emotions. Other datasets \cite{lang1997international,kim_building_2018} are built by assessing images using Self Assessment Manikin (SAM) \cite{chanel_emotion_2006}. The subjects are either hired offline \cite{lang1997international} or employed via online crowdsourcing \cite{kim_building_2018}. The VAD scores are very useful for visual storyline generation algorithms. They can be directly fed into a neural network or linear regression models to predict a final score for each image/video. However, the scales of existing datasets containing VAD information tend to be small.

\subsubsection{Aesthetics} Aesthetics scores reflect how beautiful an image is. \citeauthor{murray2012ava} \cite{murray2012ava} collected images and the corresponding aesthetics ratings from DPChallenge.com. It is a digital photography contest website, where professional photographers provide ratings on user uploaded photos. Each image has 210.13 annotations on average, with a standard deviation of 61.51. Other datasets \cite{tang_content-based_2013,kong2016photo} employ crowdsourcing workers to rate image aesthetics in the datasets. Datasets such as \cite{schwarz_will_2018,ding_intrinsic_2019} use the like rates (i.e., \#likes divided by \#views) on Flickr as the aesthetics labels. 
\citeauthor{schwarz_will_2018} \cite{schwarz_will_2018} considers the score distribution to ensure that the dataset follows a uniform distribution instead of normal distributions in other datasets \cite{murray2012ava,kong2016photo}. In addition, the Flickr-AES dataset \cite{ren2017personalized} also provides worker IDs for each annotation, which enables the personalized aesthetics score estimation. 

\subsubsection{Memorability} The memorability score reflects how many viewers can remember an image/video after a certain period of time. A more memorable video can be more helpful in persuading a viewer to purchase a product in time to come \cite{okeefePersuasionTheoryResearch2016}. A commonly-used method to collect memorability score is through visual memory games \cite{khosla_understanding_2015}. In such a game, a sequence of images are displayed to crowdsourcing workers, each of which is displayed for a short time (e.g., 500ms), with a certain short gap (800ms \cite{khosla_understanding_2015}) in-between images. The task for a viewer is to press a button whenever he/she sees an image at he/she has seen before. \citeauthor{newman2020multimodal} \cite{newman2020multimodal} considers video memorability at different replay elapse time scales. VideoMem \cite{cohendet2019videomem} provides short-term and long-term memorability annotations. They are obtained by repeating the videos after a few minutes (for short-term annotations) and 24-72 hours (for long-term annotations).

\begin{table}[]
\resizebox{1.0\linewidth}{!}{ \begin{tabular}{|l|l|l|l|l|l|l|l|}
\hline
             & Dataset         & Task           & Avg. Length (Sec) & \# video & Source                                                            & \# category & Annotation                         \\\hline
Emotion      & VideoEmotion-8 \cite{jiang2014predicting} & classification & 107         & 1,101          & Youtube                                                           & 8 \& 24     & Crowdsourcing                      \\\hline
Emotion      & YF-E6 \cite{xu2016heterogeneous}           & classification & 112         & 1,637          & Youtube and Flickr                                                & 6           & Crowdsourcing                      \\\hline
Memoribility & Memento10k \cite{newman2020multimodal}      & regression     & 3           & 10,000         & Internet                                                          & N.A.        & Crowdsourcing \\\hline
Memorability & VideoMem \cite{cohendet2019videomem}        & regression     & 7           & 10,000         & TRECVID and Hollywood-like movies & N.A.        & Crowdsourcing \\\hline
\end{tabular}}
\caption{Video-based Cognitive Property Assessment Datasets}\label{tb:3}
\end{table}

\subsection{Online Testing}
The importance of performing online testing on the generated promotional videos is two-folded. Firstly, online testing can directly reflect the impact of the generated videos on viewers' purchasing behaviours. We refer it this as \textit{video-level assessment}. Secondly, the e-commerce system administrators might want to compare two sets of videos generated by different algorithms to determine which algorithm to adopt. We refer to this as \textit{algorithm-level assessment}. Online testing can be used to evaluate storyline generation and post-production.

The most accurate evaluation method is A/B testing in a real-world e-commerce platform \cite{kohavi2007practical}. The platform can use a small percentage of the production traffic for testing. Half of the testing traffic goes into bucket A which shows videos generated by algorithm $a$. The other half goes into bucket B which shows videos generated by algorithm $b$. Actual business metrics (e.g., add-to-cart click rate, number of orders per view or GMV per view) of each bucket or video are recorded. By comparing the metrics, both video-level and algorithm-level assessments can be performed. Even though it can reflect the real-world performance of the video generation models, it is an expensive process and researchers may not have access to the A/B testing in real-world e-commerce platforms.

\begin{figure}
    \centering
    \includegraphics[width=0.6\linewidth]{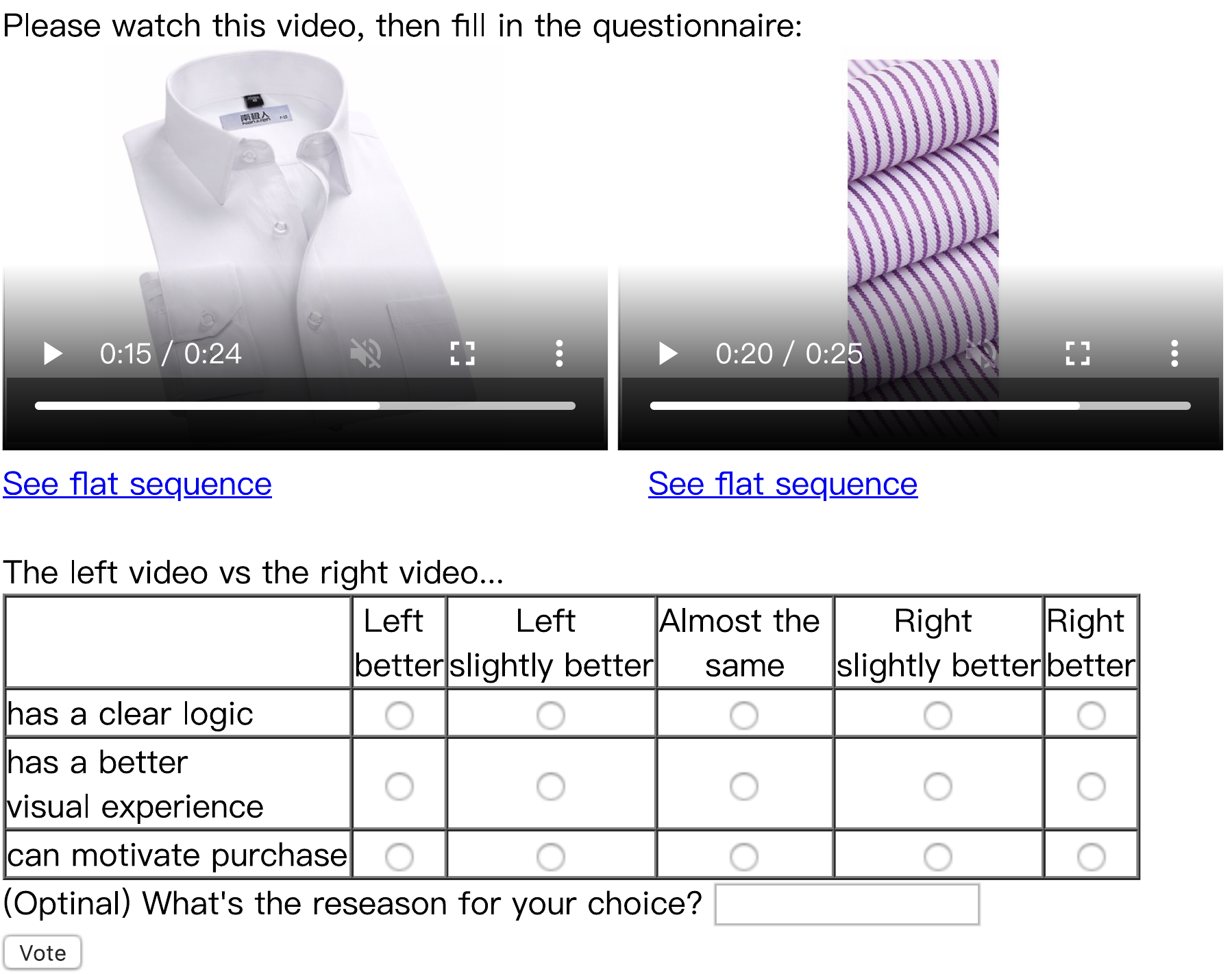}
    \caption{An example pairwise comparison task sent to mTurk workers.}
    \label{fig:screenshot}
\end{figure}

A more accessible way of performing A/B testing for researchers is through crowdsourcing platforms such as Amazon mTurk \footnote{\url{www.mturk.com/}}. This is commonly known as ``pairwise comparison'' \cite{sigurdsson2016,choi_video-story_2016,zhong2018,liu_generating_2019,liu2021enhancing}. A micro-task sent to the workers is illustrated in Figure~\ref{fig:screenshot}. Two videos are presented to each worker. The videos are about the same product, but generated by different algorithms. After watching the videos, workers are asked to provide ratings according to a 5-point Likert scale \cite{albaum1997likert} about whether the left hand side video is better than the right hand side video according to some predefined criteria (e.g., persuasiveness, attractiveness, logic flow). The performance of the video/algorithm is calculated by collecting and averaging the feedback from the workers. The pairwise comparison method might not reflect reality as accurately as A/B testing in e-commerce platforms as it is more subjective in nature \cite{aguinis2021mturk}. 

\section{Promising Future Research Directions}
Through discussions in this review, it can be observed that the building blocks for each of the major steps required for AI-empowered automatic generation of promotion videos have been developed. The current state-of-the-art techniques can be used to generate videos that can serve the business objectives of e-commerce platforms to some extent. Nevertheless, research in this interdisciplinary field is still in its early stage. Many research problems remain to be solved. In this section, we highlight promising future research directions for AIPVG.

\subsection{Visual Material Understanding}
\subsubsection{Scaling Up Proxy-based Deep Metric Learning} As discussed, a large number of classes in the training dataset leads to a huge proxy matrix. Although it is not used for inference (i.e., representation extraction), it still incurs huge memory and computational costs. The pair-based loss reduces the computational costs by using real data pairs to train the model. However, it suffers from performance loss in some cases or datasets. For example, the proxy-based loss performs better on the Person Re-identification dataset \cite{Luo_2019_CVPR_Workshops,Luo_2019_Strong_TMM} than the pair-based loss. \citeauthor{movshovitz2017no} \cite{movshovitz2017no} has enhanced proxy-based DML to only use $\frac{M}{2}$ proxies ($M$ is the number of classes), albeit at the cost of a severe drop in model accuracy. Enabling proxy-based DML training with a small number of proxies while maintaining model performance remains an open research problem.

\begin{figure}[ht]
    \centering
    \includegraphics[width=1\linewidth]{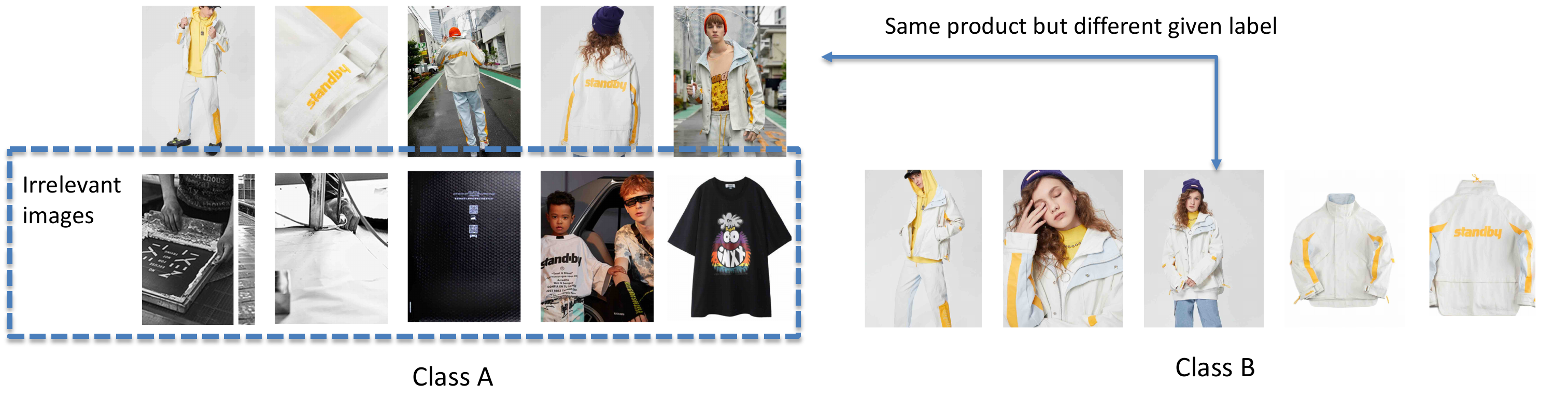}
    \caption{An example of label noise in an e-commerce product dataset for DML. Irrelevant images may appear in a class. VMs about a same product can be given different class labels due to mistakes in the data collection process. For example, multiple item IDs can refer to the same product. Thus, using item IDs in e-commerce platforms to label VMs might introduce such noises.}
    \label{fig:noise}
\end{figure}

\subsubsection{Noise Resistant Deep Metric Learning} For large scale VM datasets, label noises are inevitable. Cleaning up labels manually is an expensive process. Figure~\ref{fig:noise} illustrates the label noises that may appear in a VM dataset. Within a class, some data can be wrongly labelled. VMs about a same object might also receive different labels. Label noises decrease the performance of DML models. However, existing research works on handling noisy labels are mostly designed for classification tasks \cite{frenay2013classification,song2020learning}. 
Although efforts to enable DML to work in the presence of label noises have emerged \cite{liu2021noise}, the general approach of removing suspected noisy samples based on analyzing their average distances with clean data points still results in large accuracy drops when the noise ratio is high. DML models which are robust in the presence of label noises are still required in order to enhance the real-world applicability of VM understanding approaches.

\subsubsection{Understanding and Extending Pairwise-Comparison Models in CPA} The high accuracy of pairwise comparison models \cite{lee2019image} raises some interesting questions. Firstly, pairwise-comparison models are not commonly used in tasks like object recognition, as the state-of-the-art solutions generate predictions without explicit reference points. What is the reason behind the success of the pairwise-based methods in aesthetics estimation tasks? Secondly, given the fact that comparing a target image with all reference images is time-consuming, how to make the evaluation more efficient? Thirdly, existing approaches select the reference images based on human knowledge. Is is possible to perform this action without human intervention? Last but not least, can pairwise comparison models built for aesthetics estimation tasks improve the accuracy of emotion regression and other CPA tasks?

    
\subsection{Visual Storyline Generation}
\subsubsection{Building Datasets of E-commerce Promotional Videos} Previous methods of visual storyline generation for promotional video production heavily relies on human priors \cite{liu_generating_2019,liu2021enhancing}. However, these priors may not always lead to the optimal sequence of VMs as the application scenarios change. Data-driven approaches can be more adaptable for such situations by taking user feedbacks into account. Specifically, user feedbacks can be in the form of user ratings on the storyline, or changes in purchasing behaviours after viewing the generated videos. Such user feedback signals can be leveraged to continually train machine learning models. These models can be generative models for generating visual storylines directly, or storyline rating models for estimating a score for a given visual storyline. In the latter case, a search algorithm is also required to find suitable storylines. All these require the existence of a large-scale visual storyline dataset related to e-commerce promotional videos. Currently, no such dataset is publicly available. In order to address this issue, more indepth open collaboration between e-commerce platform operators and the AI research community is necessary. 

\subsubsection{Personalized Visual Storyline Generation} As postulated by the Persuasion Knowledge Model (PKM) \cite{friestad1994persuasion}, to maximize persuasiveness, the persuader should adapt the persuasion attempt based on knowledge about the persuadee. In a real-world e-commerce platform, it is natural that different customers have different preferences. Existing AIPVG approaches mostly generate a fixed storyline for a given product for viewers. However, this approach does not account for differences in personal preferences. Personalized visual storyline generation can be a promising research direction to enhance the persuasiveness of the resulting videos. A possible approach is to incorporate user prior information into AIPVG by leveraging personalization techniques in content-based recommender systems \cite{anand2020content,deldjoo2020recommender}. 

\subsubsection{Human Crafted Visual Storyline Templates} Human crafted visual storylines, either by experts and crowdsroucing workers, can be useful for AIPVG. Despite being one of the easiest methods to implement in real-world e-commerce platforms, they have not received much research attention so far. In order for these approaches to be leveraged to enhance AIPVG, some important research problems must be tackled first. They include: 1) how to automatically evaluate the quality of the crowdsrouced storylines for given products; 2) how to determine the suitablility of the crafted visual storylines for given products; and how to determine the level of appeal to different viewers of each given visual storyline template?

\subsection{Post-Production}
\subsubsection{End-to-End Background Music Generation} Background musics (BGMs) improve the sense of immersion for video viewers. BGM generation which considers video-music consistency is useful for AIPVG but a challenging research problem. One promising approach is to develop an end-to-end model training solution. Achieving this goal requires the following research problems to be tackled: 1) building datasets, which facilitates model training and offline evaluation (videos and soundtracks which hold Creative Common (CC) licenses are a promising source to explore); 2) model architectures, for which both accuracy and model throughput should be considered; and 3) optimization methods, which need to be custom designed to train such models effectively (e.g., adversarial training \cite{pan2019recent} could be useful).

\subsubsection{Personalization in Subtitle Generation} Informative product descriptions are helpful for the customers to make purchasing decisions. The visual contents in promotional videos may be attractive, but key product information in the images/videos might be overlooked by the viewers when viewing the videos. This could be caused by the low resolution of generated videos, the small size of the devices used to display the videos, or viewers' attention being diverted by other things when viewing the videos. One way to solve this issue is to explicitly present key information (e.g., product selling points) to the viewers. In this way, such information enhances viewers' impressions about the product. The selling points can be presented as keywords or short phrases in the form of subtitles, or as a sentence with voice-overs. Existing works in image/video captioning \cite{sharma2020image,dong2019personalized} and textual selling point mining \cite{Xu-et-al:2022IAAI} can be useful starting points. 

\subsubsection{Personalization in Visual Effect Optimization} Visual effects (VFX), such as zooming-in, zooming-out and motion blurring, bring more dynamism to videos. A wide variety of visual effects can be implemented with the help of computer vision techniques (e.g., zooming into the product in an image using the product detection and segmentation \cite{minaee2021image}). However, the most suitable types of visual effects vary across different VMs, products and viewers. 
New visual effect optimization approaches that can balance improvements in persuasion power with computational costs incurred are required.


\subsection{Evaluation}
\subsubsection{Offline Video Persuasiveness Assessment} Currently, there is no methodology for evaluating the persuasiveness of videos without involving human viewers. This limits the advancement of video-based persuasion technology since user study-based evaluation is costly and time-consuming. To address this limitation, a video persuasiveness assessment dataset is required. Such a dataset should capture the diverse backgrounds of the viewers and their behaviours when rating the perusasiveness of different types of videos so that useful viewer behaviour phenotypes can be constructed. Based on such as a dataset, persuasiveness evaluation metrics which are only related to video features need to be designed so that researchers can circumvent the need for involving human viewers when continually improving the persuasiveness of the generated videos.


\section{Conclusions}
In this paper, we provide a comprehensive survey on the artificial intelligence (AI)-empowered persuasive video generation (AIPVG) literature. We first introduce the theoretical foundations for many of such works, which is the persuasion theory.  Then, we offer a unique taxonomy of the AIPVG literature which divides it into three steps: 1) visual material understanding; 2) visual storyline generation; and 3) post-production. We analyse the design rationale, advantages and limitations of the approaches under each of these three steps. In addition, we also introduce the metrics and datasets for evaluating the performance of AIPVG approaches. Through this survey exercise, we point out promising future research directions that are worth exploring. We hope that this survey can serve as a roadmap for engineers and researchers interested in the field of AIPVG to acquire insights and build useful applications.

\section*{Acknowledgements}
This research is supported, in part, by Alibaba Group through Alibaba Innovative Research (AIR) Program and Alibaba-NTU Singapore Joint Research Institute (JRI), Nanyang Technological University, Singapore; the National Research Foundation, Singapore, under its AI Singapore Programme (AISG Award No: AISG2-RP-2020-019); Nanyang Assistant Professorship (NAP); the RIE 2020 Advanced Manufacturing and Engineering (AME) Programmatic Fund (No. A20G8b0102), Singapore. Any opinions, findings and conclusions or recommendations expressed in this material are those of the author(s) and do not reflect the views of National Research Foundation, Singapore.

\bibliographystyle{ACM-Reference-Format}
\bibliography{sample-base}

\end{document}